\documentclass[conference]{IEEEtran}
\IEEEoverridecommandlockouts
\usepackage{cite}
\usepackage{amsmath,amssymb,amsfonts}
\usepackage{mathtools}
\usepackage{algorithmic}
\usepackage{graphicx}
\usepackage{textcomp}
\usepackage{xcolor}
\usepackage{booktabs}
\usepackage{graphicx}
\usepackage{subcaption}  % for subfigures
\usepackage{pgfplots}
\pgfplotsset{compat=1.18}

\usepackage{todonotes}
\usepackage{quantikz}
\usepackage{tikz}

\begin{document}

% \title{Distributive Realization of Quantum Codes: An Intra-Node Approach\\
\title{Networked Realization of Quantum LDPC Codes\\
% \thanks{The work of S.~S. and N.~R. was partially supported by the National Science Foundation under Grant nos. 2106189 and 2540171.}
}

\author{%
   \IEEEauthorblockN{Swayangprabha Shaw and Narayanan Rengaswamy} \\
   \IEEEauthorblockA{Department of Electrical and Computer Engineering,
           University of Arizona, Tucson, AZ 85721, USA \\
           Email: \{ swayangprabha23 , narayananr \}@arizona.edu}%
  }

\maketitle

\begin{abstract}
Quantum low-density parity-check (QLDPC) codes with good parameters form the most promising candidates for low-overhead fault-tolerant quantum computing.
But a price to pay for their good parameters is the highly non-local long-range structure of their stabilizers.
While some hardware platforms can support much denser qubit connectivity than others, this necessitates the constant movement of many qubits, which introduces other challenges such as heating.
Prior work has studied the networked implementation of topological codes, where each node only holds one or a few qubits of the entire code, and demonstrated competitive performance under practical constraints such as the quality of network-provided entanglement.
However, since these codes are already geometrically local, such a networked setting might not be essential.
In this work, we propose and study the networked implementation of better QLDPC codes, specifically bivariate bicycle codes due to their similarity to surface codes and the controlled amount of long-range connections in their stabilizers.
We begin by recreating networked surface codes in \texttt{Stim}, with one code qubit allocated per node, and produce further insights on their circuit-level noise performance than was shown in prior work.
Subsequently, we extend the approach to bipartitions of bivariate bicycle codes, where we use the balanced min-cut partitioning on their combined $X$-$Z$ Tanner graph to identify the optimal split of qubits.
We exploit teleported CNOTs for those stabilizers that straddle the two nodes and vary the Bell pair fidelity that enables these gates.
Through circuit-level noise simulations on several codes decoded with BP-OSD, we provide the first insights on the networked realizations of these codes and compare their performance with monolithic implementations.
We conclude with a discussion of the advantages and disadvantages of this approach, providing an outlook for future research in such realizations of even better codes, e.g., hypergraph product and lifted product codes.
\end{abstract}


\section{Introduction}

\IEEEPARstart{F}{ault-tolerant} quantum computation is widely regarded as a necessary condition for realizing large-scale quantum advantage. 
Among the various error correction-based approaches, topological codes such as the surface code have emerged as leading candidates due to their high noise thresholds and compatibility with nearest-neighbor architectures. 
However, due to their vanishing encoding rate and the resulting large space overhead, the focus has shifted towards quantum low-density parity-check (QLDPC) codes with better code parameters.
Most existing studies on QLDPC codes have assumed a monolithic architecture, where all qubits are locally connected within a single device to realize the code and perform logical operations.
Since a price to pay for better parameters is the non-local nature of the code stabilizers, it is generally challenging to engineer such connectivity in a single processor.
% But from the quantum hardware perspective, such monolithic designs face significant engineering challenges for scaling up, motivating the exploration of networked quantum computing architectures.
This has motivated the exploration of networked quantum computing architectures, where multiple smaller quantum modules are interconnected via quantum and classical communication channels.  
Since distributed quantum computing typically refers to splitting an algorithmic task into subtasks that are executed on distinct complete processors (potentially error-corrected), we use ``networked'' to mean that a single code is realized by networking multiple nodes and each node by itself is not a complete processor.

This architecture is not merely a scaled-down replacement for a monolithic processor but it changes how fault-tolerance must be implemented. 
In particular, stabilizer measurements that were local within a single device become non-local across modules, and therefore depend on the generation, distribution, and consumption of shared entanglement.
From this perspective, the performance of networked fault-tolerant quantum computation is determined not only by local gate fidelities, but also by the rate, fidelity, and latency of inter-module entanglement resources used to mediate non-local syndrome extraction and logical operations. 
%\cite{de_Bone_2020}\cite{chandra2025distributedrealizationcolorcodes}.
Some theoretical works suggest that networked implementations of topological codes can retain the competitive error thresholds of their monolithic counterparts, despite having additional constraints such as teleported entangling operations~\cite{Nickerson:2013ecj,deBone-arxiv24,Li-pra16,Ramette-npjqi24,Hong-arxiv23,Viszlai-arxiv23,Guinn-arxiv23,chandra2025distributedrealizationcolorcodes}.

The work of Nickerson \emph{et al.}~\cite{Nickerson:2013ecj} motivates this project because they first studied the error correction threshold for a networked implementation of the surface code.
They considered an architecture where each surface code qubit is located in its own node (module), along with $3$-$4$ ancillary qubits used to create noisy Bell pairs with adjacent nodes and purify them.
% Our generalized version of this architecture for QLDPC codes is depicted in Figure~\ref{fig:SQALE_NQC}.
The links between nodes have an error rate of $10\%$, while the intra-node operations have an error rate that is swept through a range of values to determine the threshold.
They show that by creating highly purified Bell pairs between cells, fusing them into $4$-qubit GHZ states, and performing stabilizer measurements through local CNOTs between code qubits and GHZ qubits followed by classical communication to compute the measurement parity, one can obtain a threshold of about $0.75\%$ for the intra-node operations.
This was the first time that such a high threshold, close to the $\sim 1\%$ threshold in the monolithic case~\cite{Fowler_2012}, was obtained for a networked implementation of the surface code.
However, besides algorithmic improvements to the distillation protocol~\cite{Krastanov-quantum19,de_Bone_2020}, there is limited follow-up work that expands in this direction.
In general, for any good QLDPC code, it remains open to systematically study the details of its networked realization and compare that with the monolithic setting.
This is a critical gap that must be addressed to inform the development of hardware.

In this work, to setup and verify our simulation framework, we begin by investigating the networked implementation of surface codes under realistic circuit-level noise using the \texttt{Stim} simulator~\cite{Gidney_2021}.
To evaluate logical error rates, we leverage the \texttt{PyMatching} framework which uses the Sparse Blossom algorithm for decoding~\cite{Higgott_2025}.
Beyond recovering the results from~\cite{Nickerson:2013ecj}, we shed new light into the threshold for the fidelity of GHZ states in that architecture.
Since the surface code is already geometrically local, it doesn't demand a networked implementation for scalability.
But for QLDPC codes with non-local connections, the networked implementation is fundamentally attractive since one can use shared entanglement to circumvent the direct engineering of non-local connectivity.
To explore this, we extend our study to the Bivariate Bicycle (BB) QLDPC codes. 
These codes have recently attracted a lot of attention, particularly in the context of IBM’s bicycle architecture, due to their improved encoding rates and reduced qubit overhead compared to surface codes~\cite{yoder2025tourgrossmodularquantum,Bravyi_2024}. 
While prior work typically considers an entire BB code in each network module, we explore a novel approach in which a single BB code is partitioned across multiple networked modules. 
This allows us to analyze the trade-offs between locality, connectivity, and fault-tolerance in networked QLDPC code implementations.
Note that while prior work has also investigated teleported CNOTs for BB codes~\cite{Berthusen-prxq25}, there the code still lies in one layer of a monolithic chip and a separate layer is used primarily to create long-range entanglement.
In the simplest nontrivial networked setting of a BB code split across two nodes, we show that a $\approx 1\%$ threshold exists for local gate noise, and that Bell pairs of fidelity at least $99\%$ are necessary to match the monolithic architecture.

% Through these investigations, our goal is to bridge the gap between theoretical proposals and practical implementations of networked quantum error correction, providing insights into the feasibility and performance of scalable distributed quantum computing systems.

The paper is organized as follows.
Section~\ref{sec:background} reviews the essential technical background on stabilizer codes, CSS codes, QLDPC codes, surface codes, and BB codes.
Section~\ref{sec:architecture} begins with the networked architecture for surface codes and then introduces its extension to BB codes via graph partitioning and teleported CNOTs.
Section~\ref{sec:sim_setup} provides the key details of the simulation setup in \texttt{Stim} spanning circuit construction, measurement scheduling, detector construction, decoding, and noise model.
Section~\ref{sec:results} discusses the simulation results for surface codes and BB codes in detail, highlighting the key factors in a networked architecture.
Section~\ref{sec:conclusion} summarizes and concludes the paper with an outlook for future work.

\section{Technical Background}
\label{sec:background}

\subsection{Stabilizer Codes}

This is a class of quantum error-correcting codes for protecting quantum information against environmental noise and operational errors in quantum computing and quantum communication.
These codes are subspaces defined based on the Pauli group, which comprises of tensor products of Pauli matrices ($I$, $X$, $Y$, $Z$) up to a global quaternary phase $\{ \imath^\kappa; \kappa = 0,1,2,3 \}$, where $\imath = \sqrt{-1}$.
A stabilizer code is defined as the common $+1$-eigenspace of a set of commuting Pauli operators known as the stabilizer group. 
% This group is generated by independent and commuting elements. 
If there are $n-k$ independent generators for this group on $n$ qubits, then the resulting code has dimension $2^k$.
% A stabilizer state is the state that remains invariant under the action of these group elements. 
Error correction is achieved by frequently measuring these stabilizer generators; the outcomes $\pm 1$, often referred to as the error syndrome, indicate whether an error has occurred.
The syndrome string is passed on to a decoder to identify the most likely error, given a statistical model for the error channel.

The capability of stabilizer codes to correct errors depends on their distance, defined as the minimum weight (number of non-identity elements) of an undetectable (Pauli) error, i.e., any operator in the Pauli group that commutes with all elements of the stabilizer but is not in the stabilizer itself. 
A code with distance $d$ can correct up to $\lfloor (d-1)/2 \rfloor$ errors (ignoring degenerate errors), making it robust against a significant amount of quantum noise.
An $[\![ n,k,d ]\!]$ code encodes $k$ logical qubits into $n$ physical qubits and has distance $d$.

\subsection{Calderbank-Shor-Steane (CSS) Codes}

This subclass of stabilizer codes is constructed from two classical linear codes that satisfy a specific orthogonality condition~\cite{PhysRevA.54.1098,Steane1996MultipleparticleIA}. 
In a CSS code, the stabilizer group can be generated using two distinct subgroups: one consisting entirely of tensor products of Pauli-$X$ operators and the other of Pauli-$Z$ operators. 
This separation is particularly advantageous because it allows $X$-type (bit-flip) and $Z$-type (phase-flip) errors to be detected and corrected independently (in the absence of error correlations), considerably simplifying both the syndrome extraction and decoding procedures. 
% The surface code, which forms the basis of the topological error correction scheme employed 
% in this work, is a prominent example of a CSS code.  — a property that makes it especially well suited for 
% physically realistic, nearest-neighbour quantum architectures.

\emph{Quantum low-density parity-check (QLDPC) codes} are families of stabilizer codes with common properties and growing size $n$ where each stabilizer only involves $O(1)$ qubits and each qubit is only involved in $O(1)$ stabilizers, independent of $n$.
Since these codes only require local connectivity for syndrome measurements, albeit not necessarily geometrically local, they are particularly suited well for scalable physical implementations.
A code is said to be a QLDPC code if it is a member of a QLDPC family of codes.
Such a code is represented by its bipartite Tanner graph and decoded using iterative message passing algorithms on the graph, such as belief propagation (BP).
While QLDPC codes can be general stabilizer codes, they are most often CSS codes.

\subsubsection{Surface Codes}

This is a prominent family of CSS QLDPC codes that has attracted significant attention in the field, owing to their robustness against local errors and their natural compatibility with two-dimensional qubit architectures~\cite{Kitaev_2003,Fowler_2012}. 
A planar (unrotated) surface code of distance $d$ is characterized by the parameters $[\![n, k, d]\!]$, where $n = 2d^2 - 2d + 1$ and $k = 1$. 
%is the total number of physical qubits (comprising both data and ancilla qubits), $k = 1$ is the number of logical qubits encoded, and $d$ is the code distance — the minimum number of physical errors required to cause an undetectable logical failure. 
% A surface code of distance $d$ can correct up to $\lfloor (d-1)/2 \rfloor$ arbitrary errors on the data qubits. 
The construction of a surface code involves arranging qubits on a $d \times d$ planar lattice, where data qubits reside on the edges while syndrome ancilla qubits are positioned on the faces and vertices of the lattice. 
This ensures that all stabilizer measurements remain geometrically local, requiring only nearest-neighbor interactions, which is a highly favorable property for physical implementation.

The stabilizers of the surface code partition naturally into two types: 
(1) plaquette operators, defined as the product of Pauli-$Z$ operators around each face, and 
(2) star (or vertex) operators, defined as the product of Pauli-$X$ operators around each vertex. 
These operators are typically of weight four or fewer, and their eigenvalues, obtained through projective measurement, yield a syndrome that indicates whether an error has occurred, without disturbing the encoded quantum information. 
% The clean separation between $X$-type and $Z$-type stabilizers, inherited from the underlying CSS structure, allows bit-flip and phase-flip errors to be detected and corrected independently, considerably simplifying the decoding process.
A key advantage of the surface code is its comparatively high fault-tolerance threshold, estimated to be in the vicinity of $1\%$ depending on the specific implementation and noise model~\cite{Fowler_2012}. 
However, this robustness comes at a considerable cost in physical resources. 
Since the surface code encodes only $k = 1$ logical qubit regardless of code distance, the number of physical qubits scales as $O(d^2)$ per logical qubit.
This space overhead leads to thousands to millions of physical qubits for practically useful logical error rates~\cite{Fowler_2012,Gidney_2021_RSA}.
This substantial overhead motivates the exploration of more resource-efficient 
architectures with robust codes. %, such as the networked approach with Bivariate Bicycle code~\cite{Bravyi_2024,yoder2025tourgrossmodularquantum}.
% This threshold is substantially higher than that of many other quantum 
% error-correcting codes, making the surface code a leading 
% candidate for scalable, fault-tolerant quantum computing in 
% settings where physical qubits are inevitably subject to noise.

\subsubsection{Bivariate Bicycle (BB) Codes}

\begin{figure}[!ht]
    \centering
    \vspace*{-10pt}
    % Subfigure (a): Toric layout — top-left region
    \begin{subfigure}[b]{0.4\textwidth}
        \centering
        \includegraphics[trim={0cm 10.5cm 14cm 0cm}, clip, width=\textwidth]{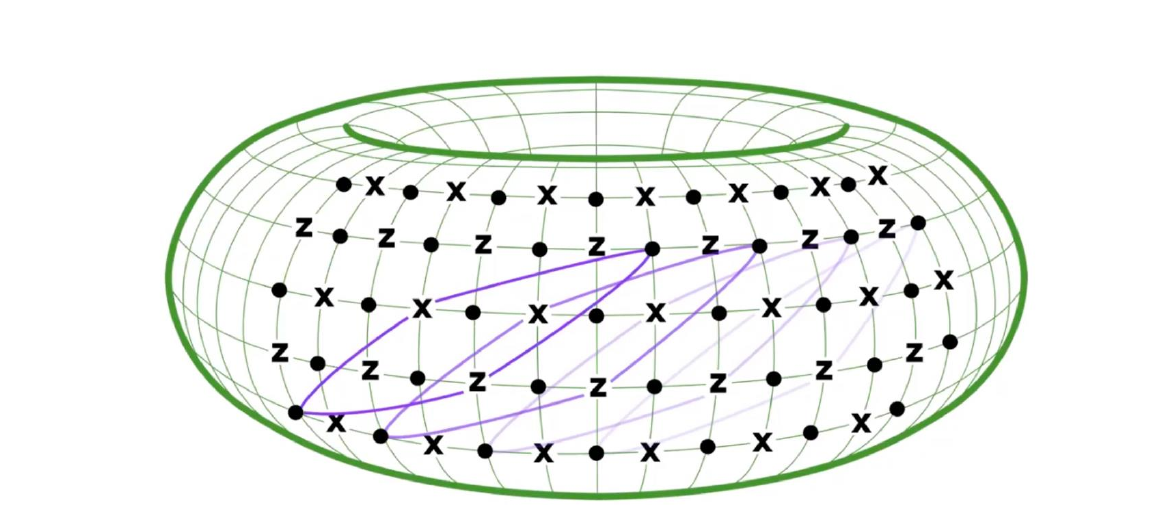}
        \caption{Geometric representation of the BB code on a torus, where data and 
    ancilla qubits are placed on a periodic 2D lattice.}
        \label{fig:bb_toric}
    \end{subfigure}
    \hfill
    % Subfigure (b): Tanner graph — top-right region
    \begin{subfigure}[b]{0.4\textwidth}
        \centering
        \includegraphics[trim={17.5cm 1.2cm 1cm 2cm}, clip, width=\textwidth]{figures/bb_code.pdf}
        \caption{Tanner graph of the code, showing data–check qubit connectivity and how the left-right data qubit is assigned. The stabilizer 
    generators are built from the polynomial pair $A = A_1 + A_2 + A_3$ 
    and $B = B_1 + B_2 + B_3$.}
        \label{fig:bb_tanner}
    \end{subfigure}
    \hfill
    % Subfigure (c): Depth-8 scheduling — bottom region
    \begin{subfigure}[b]{0.5\textwidth}
        \centering
        \includegraphics[trim={3cm 1.2cm 16.5cm 9.5cm}, clip, width=\textwidth]{figures/bb_code.pdf}
        \caption{Depth-8 CNOT scheduling used for fault-tolerant syndrome 
    measurement, specifying the order of CNOT gates between data and ancilla qubits over eight time steps.}
        \label{fig:bb_schedule}
    \end{subfigure}
    
    \caption{Bivariate bicycle (BB) code~\cite{Bravyi_2024}: 
    (a) toric representation, 
    (b) Tanner graph, and 
    (c) depth-8 measurement schedule.}
    \label{fig:bb_code}
    \vspace*{-5pt}
\end{figure}

A BB code is constructed from two commuting cyclic-shift operators,
\(
x = S_{\ell}\otimes I_m
\)
and
\(
y = I_{\ell}\otimes S_m
\),
where \(S_{\ell}\) and \(S_m\) are cyclic shift matrices and therefore satisfy \(x^{\ell}=y^{m}=I_{\ell m}\) and \(xy=yx\).
From a group-theoretic perspective, it is a CSS QLDPC code constructed from sparse bivariate polynomials over the ring
% \begin{equation}
$R = \mathbb{F}_2[x,y]/(x^{l}-1,\; y^{m}-1)$.
% \end{equation}.
These polynomials can identified with two binary matrices
\(
A=A_1+A_2+A_3
\)
and
\(
B=B_1+B_2+B_3
\),
where each \(A_i\) and \(B_j\) is chosen as a power of \(x\) or \(y\) (see Fig.~\ref{fig:bb_code}). 
The resulting BB code %, denoted \(QC(A,B)\), 
has code length \(n=2\ell m\) and CSS check matrices
\(
H_X=[A \mid B]
\)
and
\(
H_Z=[B^T \mid A^T]
\).
Since \(A\) and \(B\) commute, one has
\(
H_X H_Z^T = AB + BA = 0 \pmod 2
\),
so the \(X\)- and \(Z\)-type stabilizer checks commute as required. 
This code was first introduced in \cite{PhysRevA.88.012311,Panteleev_2021} and then was instantiated with the degree three polynomials in~\cite{Bravyi_2024}.
In this construction, each row and column of \(A\) and \(B\) has weight three, which gives rise to weight-$6$ stabilizer checks and degree-$3$ qubits. %a sparse Tanner graph, making BB codes an important family of quantum LDPC codes with both high encoding efficiency and structured syndrome extraction.
BB codes are attractive because they combine sparse parity checks with higher encoding efficiency than the surface code at comparable protection levels \cite{Bravyi_2024}. 
This makes them promising candidates for fault-tolerant quantum computation with modular quantum architectures~\cite{yoder2025tourgrossmodularquantum}, although their stabilizers require some nonlocal connectivity in a hardware realization.

\section{Networked Architecture}
\label{sec:architecture}

Our approach proceeds in two stages. 
We first establish a validated baseline by reproducing the networked surface code protocol of Nickerson \emph{et al.}~\cite{Nickerson:2013ecj} at the circuit level using the \texttt{Stim} simulator. 
We also share some new insights on that protocol and its performance.
Then, we extend this framework to the more complex setting of BB codes, where the nonlocal structure of the stabilizers makes a networked implementation not merely convenient but arguably necessary.

\subsection{Networked Surface Codes}

\begin{figure}[!t]
    \centering
    \includegraphics[trim={2cm 1cm 3cm 2cm}, clip, width=0.4\textwidth]{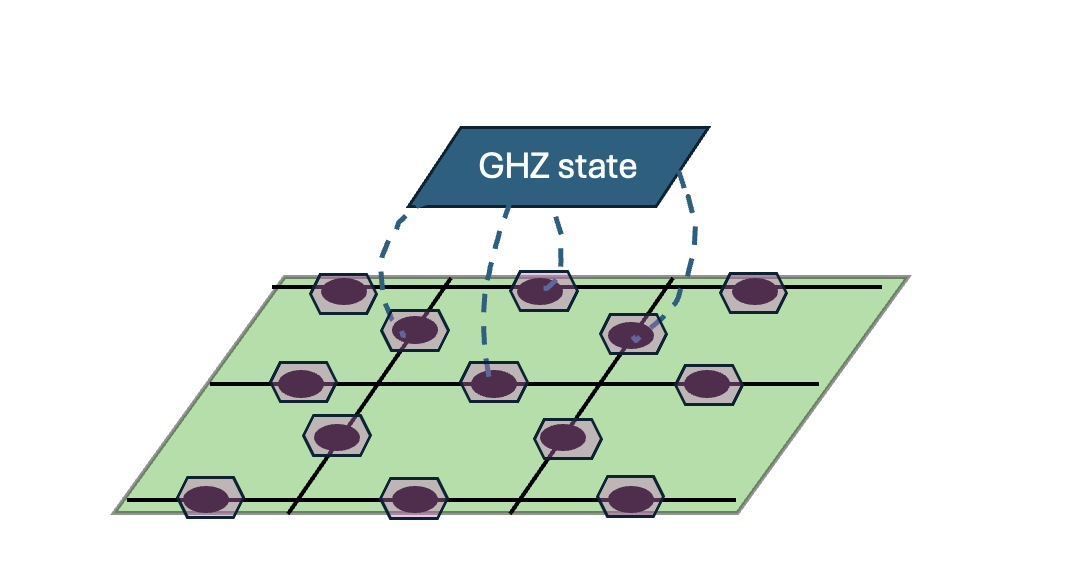}
    \caption{Schematic of networked surface code architecture. The hexagonal regions represent individual processors, each containing one data qubit. A GHZ state is utilized for syndrome extraction, as illustrated by the dashed lines connecting the GHZ state to specific data qubits within the grid structure.}
    \label{fig:ghz_syndrome_extraction}
    \vspace*{-5pt}
\end{figure}

In the networked surface code setting of~\cite{Nickerson:2013ecj}, each processor contains four qubits, one being a surface code data qubit and the other being ancilla qubits to generate shared entangled states between processors. 
%(forming one type of stabilizer) and are connected by noisy entangled links.
Since a syndrome measurement now involves four qubits on four nodes, the ancilla qubits of those nodes are used sequentially to generate Bell pairs, purify them, and fuse to form a shared $4$-qubit GHZ state shared among those nodes (see Fig.~\ref{fig:ghz_syndrome_extraction}).
Then, a stabilizer can be measured through local CNOTs between each data qubit and the respective GHZ qubit, followed by measurements and classical communication of the results.
% Stabilizer measurements are performed using the GHZ-based protocol introduced by Nickerson et al\cite{Nickerson:2013ecj}.
The original analysis of this protocol employed a  superoperator noise model, which consolidates the net effect of all imperfections into an effective noise channel acting on the data qubits.
In contrast, our \texttt{Stim}-based implementation models noise at the circuit level, meaning every gate, reset, and measurement operation in the protocol is subject to stochastic depolarizing errors: % with a given probability: Specifically, we model:
\begin{itemize}
    
    \item \textbf{GHZ noise:} GHZ states are generated with infidelity $p_{err}$, capturing the imperfections of the quantum link. We use the relation between Bell pair fidelity and GHZ fidelity (after fusion) plotted as ``Stringent'' in~\cite[Fig. 7]{de_Bone_2020} to estimate the $p_{err}$ corresponding to the $10\%$ link error rate assumed in~\cite{Nickerson:2013ecj}, i.e. $p_{err} \approx 10^{-3}$.
    
    \item \textbf{Gate noise:} Each two-qubit gate (controlled-$Z$ or controlled-$X$) is followed by a two-qubit depolarizing channel with error probability $p_g=p$.
    
    \item \textbf{Measurement noise:} Each single-qubit measurement outcome is flipped with probability $p_m=p$.
    
    \item \textbf{Reset noise:} Each qubit reset operation is followed by a bit-flip error with probability $p_r=p$.
    
\end{itemize}
The performance of this networked realization of surface codes will be discussed later along with other results.

\subsection{Networked BB Codes}

Having validated the simulation framework on surface codes, we turn to the central contribution of this work, which is the networked realization of BB codes.
However, unlike the surface code setting where the extreme networked scenario of one data qubit per node was considered in~\cite{Nickerson:2013ecj}, here we consider the least networked scenario of exactly two nodes, i.e., the BB code qubits are split across two nodes.
This setup can be naturally extended to multiple nodes as appropriate.

\subsubsection{Bipartitioning via Balanced Min-Cut}

\begin{figure}[!t]
    \centering
    \includegraphics[trim={3cm 3.8cm 10cm 2cm}, clip, 
    width=0.5\textwidth]{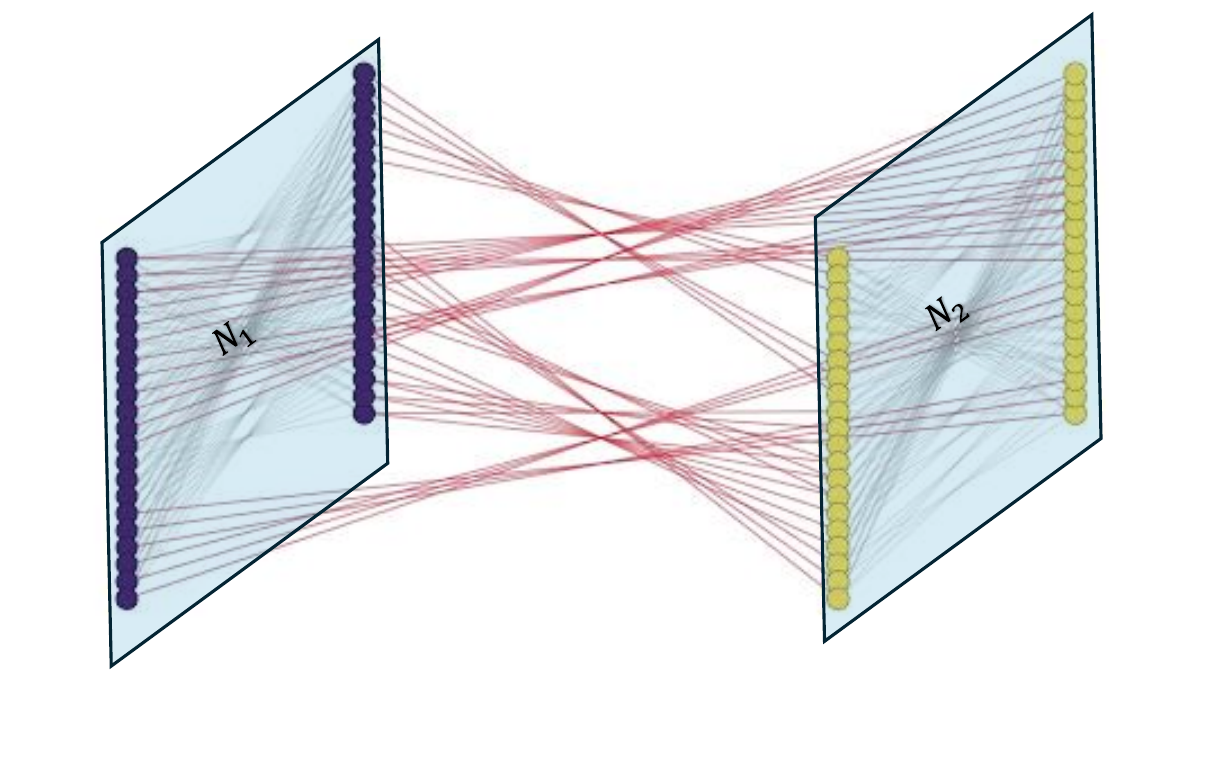}
    \caption{Balanced graph partitioning of the combined $X$-$Z$ Tanner graph across two nodes, $N_1$ and $N_2$. The red lines represent bridge edges connecting check qubits and data qubits distributed across the two nodes. We performed balanced graph partitioning on this combined $X$-$Z$ Tanner graph using \texttt{pymetis}~\cite{Karypis_1998}. This assigns every data qubit and every check qubits to one of the two network partitions.}
    \label{fig:tanner_graph_partitioning}
    \vspace*{-10pt}
\end{figure}

We construct a combined $X$-$Z$ Tanner graph of the code, where data qubits and both $X$- and $Z$-type check nodes are represented, and edges connect each check to the data qubits in its support. 
Then, we apply a balanced min-cut algorithm to this graph in order to partition the data qubits into two groups of approximately equal size, while minimizing the number of stabilizer checks that straddle the two partitions (see Fig.~\ref{fig:tanner_graph_partitioning}).
This partitioning criterion is motivated by a practical consideration---every stabilizer that spans both nodes requires at least one inter-node entangling operation, which is significantly noisier and slower than a local gate. 
Therefore, minimizing the number of such straddling stabilizers directly reduces the noise overhead introduced by the network.
The balanced constraint ensures that neither node is disproportionately loaded, which is important to maintaining synchronous stabilizer measurement cycles across the two halves of the code.

\subsubsection{Teleported CNOTs for Inter-Node Stabilizers}

For stabilizers whose support is entirely contained within a single node, syndrome extraction proceeds using standard local circuits. 
For stabilizers that straddle the two nodes, however, we employ teleported CNOT gates to implement the required entangling operations across the network link.
A teleported CNOT consumes one pre-shared Bell pair between the two nodes and requires local (noisy) operations and classical communication to complete. 
The fidelity of this gate is therefore directly governed by the quality of the shared Bell pair. 
We parameterize this quality by a Bell pair fidelity $F_{\text{Bell}}$, which we vary across our simulations to understand its impact on the overall code performance. 
In the ideal case ($F_{\text{Bell}} = 1$), the teleported CNOT is equivalent to a perfect local CNOT. 
As $F_{\text{Bell}}$ decreases, additional depolarizing noise is introduced on the data qubits involved in the cross-node stabilizer.

% For BB codes, the irregular and variable-weight structure of the stabilizers makes teleported CNOTs a more flexible and general-purpose mechanism for cross-processor entangling operations.

\subsubsection{Circuit-Level Noise Model}

As with the surface code simulations, all operations in the BB code circuits are subject to circuit-level depolarizing noise in \texttt{Stim}. Specifically:
\begin{itemize}
    \item Local single- and two-qubit gates are followed by depolarizing noise with probability $p$.
    \item Measurements and resets are subject to bit-flip errors with probability $p$.
    \item Teleported CNOTs inherit noise both from the local operations involved and from the fidelity $F_{\text{Bell}}$ of the consumed Bell pair. The effective noise on a teleported CNOT is therefore strictly greater than that of a local gate, and this asymmetry is faithfully captured in \texttt{Stim}.
    \item Idle qubits accumulate depolarizing noise at a rate $p$ per time step, accounting for decoherence during inactivity.
\end{itemize}

Decoding for the BB codes is performed using belief propagation with ordered statistics post-processing (BP-OSD), which is better suited to the non-local structure of QLDPC codes than the \texttt{PyMatching} decoder used for surface codes. 
We simulate several BB codes of varying parameters $[\![n, k, d]\!]$ and, for each, compare the performance in two settings:
\begin{itemize}
    \item monolithic implementation where all qubits reside on a single processor, and
    \item a networked implementation using our bipartitioning and teleported CNOT scheme.
\end{itemize}

\section{Simulation Setup}
\label{sec:sim_setup}

In this work, we conduct fault-tolerant memory experiments on both surface codes and BB codes using the \texttt{Stim} stabilizer-circuit simulator~\cite{Gidney_2021}.
The workflow was organized into modular stages for circuit generation, measurement scheduling, noise insertion, detector construction, and decoding.

\paragraph{Circuit generation}
A generator function, \texttt{build\_stim\_text}, was used to emit valid \texttt{Stim} circuit descriptions. 
For the surface code simulations, the circuit was constructed from the corresponding lattice geometry and parity-check structure, with binary parity-check matrices $H_X$ and $H_Z$ specifying the participation of each data qubit in the $X$- and $Z$-type stabilizers. 
For the BB code simulations, the corresponding matrices $H_X$ and $H_Z$ were used directly to define the connectivity between data qubits and check ancillas in the \texttt{Stim} circuit.
The generated circuit includes:
\begin{itemize}
    \item initialization---ideal projection of reset qubits into the code space,
    \item bare-ancilla noisy syndrome measurement rounds with detectors for each check,
    \item \texttt{OBSERVABLE\_INCLUDE} instructions for the logical $X$ or $Z$ observables,
    \item scheduled two-qubit gates according to Fig.~\ref{fig:bb_code}c implementing syndrome extraction,
    \item GHZ-based ancilla preparation for the networked surface code setting,
    \item Bell-pair bridge gadgets for the partitioned BB code setting, and
    \item circuit-level noise through \texttt{DEPOLARIZE1} and \texttt{DEPOLARIZE2}.
\end{itemize}

For the BB codes, logical observables were obtained using a symplectic Gram-Schmidt orthogonalization procedure~\cite{Wilde_2009}. %SGSOP-based construction.

\paragraph{Measurement scheduling}
\begin{itemize}
    
    \item For the surface code simulations, we used an even-odd tick-based scheduling rule to suppress hook-error propagation by temporally separating overlapping stabilizer interactions. The $X$-type and $Z$-type stabilizers were activated in alternating layers according to the parity of their indices.
    
    \item For the partitioned BB code simulations, we used the depth-8 syndrome-extraction cycle of~\cite{Bravyi_2024} for all local check--data interactions. To obtain the bipartition, we formed the combined Tanner graph associated with
    \(
    \begin{bsmallmatrix}
    H_X \\
    H_Z
    \end{bsmallmatrix},
    \)
    and partitioned this graph into two blocks using \texttt{pymetis}~\cite{Karypis_1998}. If a scheduled BB interaction connected a data qubit and check ancilla in the same partition, the original local gate from the depth-8 schedule was applied. If the interaction crossed the partition boundary, it was replaced by a Bell-pair-assisted teleported CNOT.
\end{itemize}

\paragraph{Noise insertion}
All simulations used a circuit-level noise model. 
For the partitioned BB code setting, the local noise model for idle qubits, measurements, single-qubit gates, and CNOT operations was kept fixed, while Bell-pair noise was modeled independently through a two-qubit depolarizing channel with error probability $p_{\mathrm{Bell}}$. 
Monte Carlo sampling was performed using \texttt{sinter} with 4 workers. 
Each task was run until either the maximum shot count ($10^6$) or the maximum logical-failure count ($10^3$) was reached.

\paragraph{Detector construction}
Detector instructions were inserted to track changes in syndrome outcomes between successive rounds.
In particular, detector events were formed from appropriate parity comparisons between ancilla outcomes in adjacent rounds, including the transition from the initial round to the first noisy round and from the final noisy round to the closing data readout. 
This allows the detector error model to capture faults propagating across the full measurement history.

\paragraph{Decoding}
The detector error model is decoded differently for the surface code and BB  code cases:
\vspace{0.8em}
\noindent\hspace*{2em}%

\noindent\begin{minipage}{0.7\columnwidth}
\ttfamily\footnotesize
dem = circuit.detector\_error\_model()\\
sampler = dem.compile\_detector\_sampler()\\
det\_events, obs\_flips = sampler.sample(\\
\hspace*{1.5em}shots=1000000, separate\_observables=True\\
)\\[0.5em]
\# Surface code\\
matcher\_surface =\\
pymatching.Matching.from\_detector\_error\_model(dem)\\
predictions\_surface =\\
matcher\_surface.decode\_batch(det\_events)\\[0.5em]
\# BB code\\
chk\_matrix, log\_matrix, prior\_list=dem\_matrices(dem)\\
decoder\_bb = BpOsdDecoder(chk\_matrix,error\_channel=prior\_list)\\
\end{minipage}

Here, \texttt{chk\_matrix} is the binary detector-check matrix derived from the detector error model, with rows corresponding to detectors and columns corresponding to elementary fault mechanisms.
The matrix \texttt{log\_matrix} records which fault mechanisms flip which logical observables. 
The list \texttt{prior\_list} contains the corresponding prior fault probabilities.
Logical error rates were estimated by comparing the predicted logical observable flips against the sampled observable flips.

\section{Results and Discussion}
\label{sec:results}

\subsection{Networked Realization of Surface Codes}

The foundational work of Nickerson \emph{et al.}~\cite{Nickerson:2013ecj} established critical error thresholds for networked quantum computing through the use of superoperator noise models.
This approach consolidates the complex noise characteristics of the full stabilizer protocol into effective mathematical maps acting on the data qubits. 
While this level of abstraction proved both analytically powerful and computationally efficient, it necessarily obscures the fine-grained dynamics of how errors arise, propagate, and correlate at the circuit level. 
This raises a natural question whether we can reproduce and validate these theoretical estimates using a more physically faithful, circuit-level noise model.

To begin answering the questions, we first sought to understand the impact of noise in the GHZ state preparation, which is the primary entangled resource state used to do syndrome measurement in the architecture, on the performance of the surface code.
Note that this was not analyzed explicitly in~\cite{Nickerson:2013ecj}.
To  clearly isolate this contribution, we deliberately switched off all errors on the data qubits and the circuit-level operations, leaving GHZ preparation noise as the sole source of error in the system. 
With this simplified noise environment in place, we performed a systematic sweep over a range of GHZ 
noise rates within the \texttt{Stim} setup, looking for a threshold.

\begin{figure}[!t]
    \centering
    \includegraphics[trim={0cm 0cm 0cm 0.9cm}, clip, 
    width=0.45\textwidth]{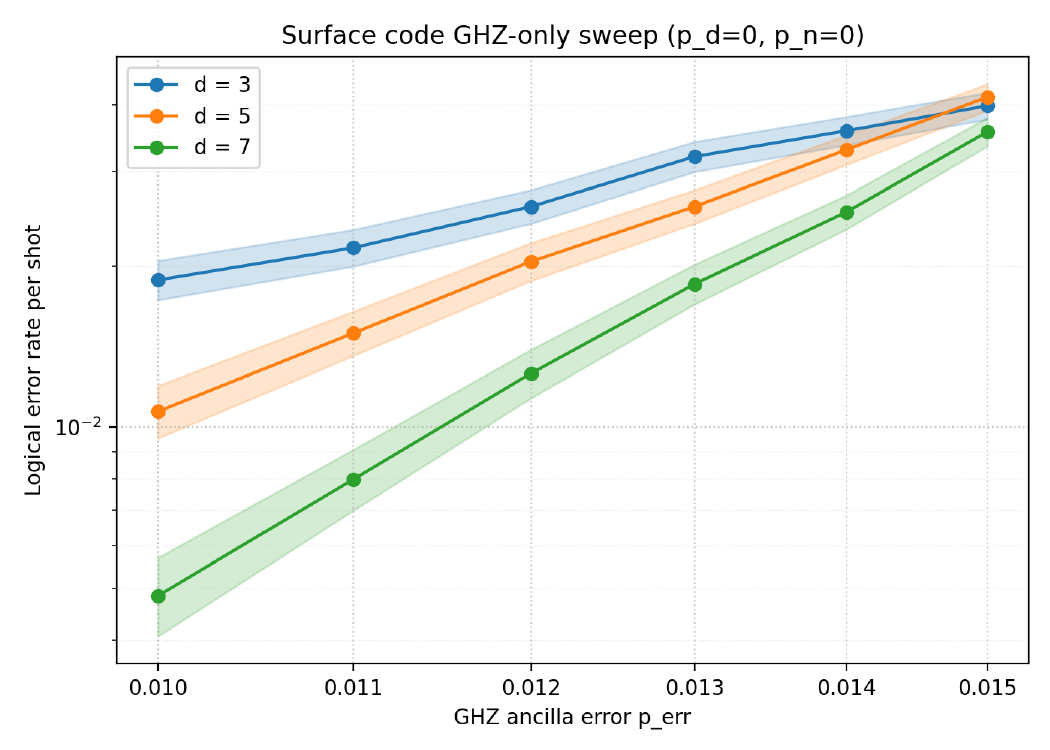}
    \caption{Surface code performance with varying GHZ noise rate 
    ($p=0$). A clear threshold is visible, below which logical 
    error rates decrease with increasing code distance.}
    \label{fig:ghzsweep}
    \vspace*{-10pt}
\end{figure}

As shown in Fig.~\ref{fig:ghzsweep}, this sweep revealed a well-defined threshold in the GHZ noise.
Below $p_{err} \approx 1.5\%$, the logical error rate falls steadily with increasing code distance, confirming that the GHZ preparation noise alone does not compromise fault tolerance provided it remains sufficiently low. 
This result offered us both a concrete noise budget for the GHZ preparation stage and, crucially, a validated sub-threshold operating point to carry forward into a more complete simulation.

Building on this foundation, we introduced circuit-level faults at the level of individual gate operations, qubit resets, and measurements, each occurring with a given probability, while fixing the GHZ noise rate at the sub-threshold value of $p_{\text{GHZ}}=0.002$.
Rather than relying on the pre-computed noise channels of the original superoperator formalism, this approach models the explicit propagation and correlation of errors as they unfold throughout the full execution of the circuit, providing a considerably more granular and physically faithful representation of the noise environment that a real quantum device would encounter.

\begin{figure}[!t]
    \centering
    \includegraphics[trim={0cm 0cm 0cm 0.9cm}, clip, 
    width=0.45\textwidth]{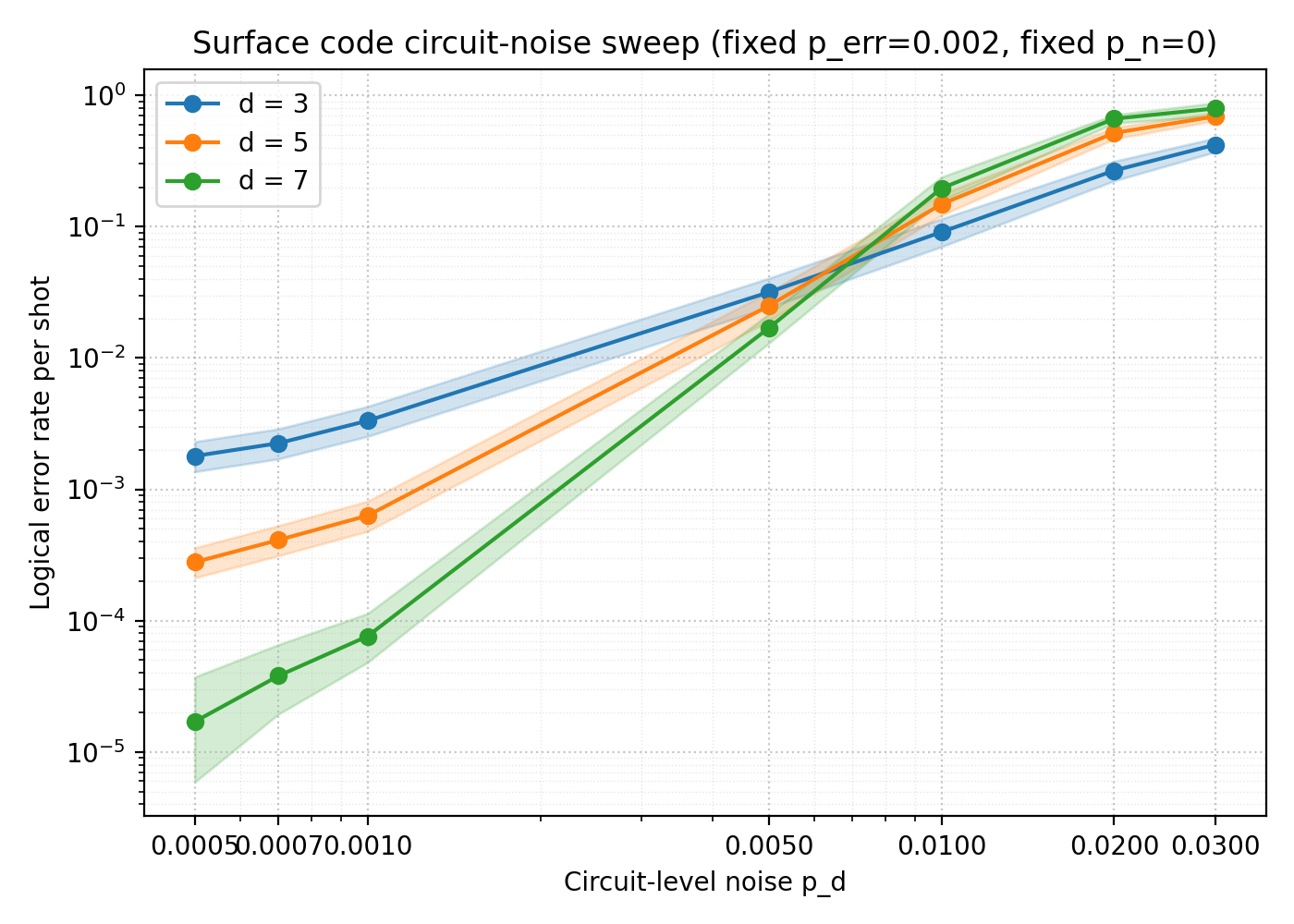}
    \caption{Surface code performance with noisy GHZ 
    ($p_{\text{GHZ}}=0.002$) under full circuit-level noise. This reproduces the exact performance and $\approx 7.5\%$ threshold estimated in~\cite{Nickerson:2013ecj}.}
    \label{fig:surface_code_ghz_0.002}
    \vspace*{-10pt}
\end{figure}

Despite this substantially more detailed treatment of noise, our simulations yield logical error rates and a threshold that are in close agreement with the original superoperator-based results of Nickerson \emph{et al.}, as shown in Fig.~\ref{fig:surface_code_ghz_0.002}. 
This correspondence is significant. 
It demonstrates that the theoretical estimates derived from the superoperator formalism are not merely convenient approximations, they remain predictive and robust even when subjected to the realistic, circuit-level fault dynamics of a physical quantum architecture.

Taken together, these results accomplish two things. 
First, the GHZ noise sweep provides a clear and practical noise budget for the network component of the protocol, offering concrete guidance on the fidelity requirements that the GHZ preparation stage must meet for fault-tolerant operation. 
Second, the full circuit-level simulation constitutes a rigorous cross-validation of the Nickerson protocol, reinforcing confidence that its theoretical framework remains a reliable guide for the design and engineering of networked architectures.

\subsection{Networked Realization of BB Codes}

For each BB code, we begin with the CSS parity-check matrices $H_X$ and $H_Z$. 
To define the bipartition used in the networked setting, we form the combined check matrix
\(
H_{XZ}=\begin{bsmallmatrix}H_X\\ H_Z\end{bsmallmatrix},
\)
which is equivalent to building a single Tanner graph containing all data-qubit vertices together with both the $X$-type and $Z$-type check vertices. In other words, vertically stacking $H_X$ and $H_Z$ does not change the data-qubit set; it combines the two check layers into one graph so that the partitioning procedure can see all check-data connections at once.
We performed balanced graph partitioning on this combined $X$-$Z$ Tanner graph using \texttt{pymetis}, a Python interface to the METIS graph partitioning library~\cite{Karypis_1998}.
This assigns every data qubit and every check node to one of two network partitions. The same partition is then used when constructing the depth-8 BB syndrome-extraction circuit.

For each scheduled check-data interaction, if the data qubit and the corresponding check ancilla lie in the same partition, the interaction is implemented as the original local CNOT from the BB schedule. If they lie in different partitions, the local CNOT is replaced by a Bell-pair-assisted bridge gadget. Therefore, the number of cut Tanner-graph edges produced by the partition is exactly the number of cross-partition interactions that must be supplied by the network.

\begin{table}[t]
\centering
\caption{Partition statistics for the $[[72,12,6]]$ BB code on the combined $X$-$Z$ Tanner graph.}
\label{tab:bb72_partition_stats}
\resizebox{\columnwidth}{!}{%
\begin{tabular}{lccc}
\hline
Quantity & X stabs & Z stabs & Total \\
\hline
Checks & 36 & 36 & 72 \\
Tanner edges & 216 & 216 & 432 \\
Bridge edges ($E_{\mathrm{cut}}$) & 51 & 53 & 104 \\
Local edges & 165 & 163 & 328 \\
Cross-partition stabilizers & 29 & 28 & 57 \\
%Spanning stabilizers & 29 & 28 & 57 \\
\hline
\end{tabular}%
}\end{table}

\begin{table}[t]
\caption{Partition statistics and Bell-pair demand for several BB codes. The subscripts ``1'' and ``2'' denote two nodes.}
\label{tab:bb_partition_summary}
\resizebox{\columnwidth}{!}{%
\begin{tabular}{lcc|cc|cc|ccc}
\hline
 & \multicolumn{2}{c|}{Data} & \multicolumn{2}{c|}{X stabs} & \multicolumn{2}{c|}{Z stabs} & \multicolumn{3}{c}{Bell-pair} \\
Code & $D_1$ & $D_2$ & $X_1$ & $X_2$ & $Z_1$ & $Z_2$ & Demand\\
% & Cuts \\
% & /cycle & /shot \\
\hline
$[[72,12,6]]$   & 35 & 37 & 17 & 19 & 20 & 16 & 104 \\
% & 104 & 624 \\
$[[90,8,10]]$   & 45 & 45 & 24 & 21 & 21 & 24 & 102 \\
% & 102 & 1020 \\
$[[144,12,12]]$ & 72 & 72 & 36 & 36 & 36 & 36 & 124 \\
% & 124 & 1488 \\
\hline
\end{tabular}%
}
\end{table}

\begin{figure*}[!t]
    \centering
    %\resizebox{0.5\textwidth}{!}{%
    \scalebox{0.8}{%
        \input{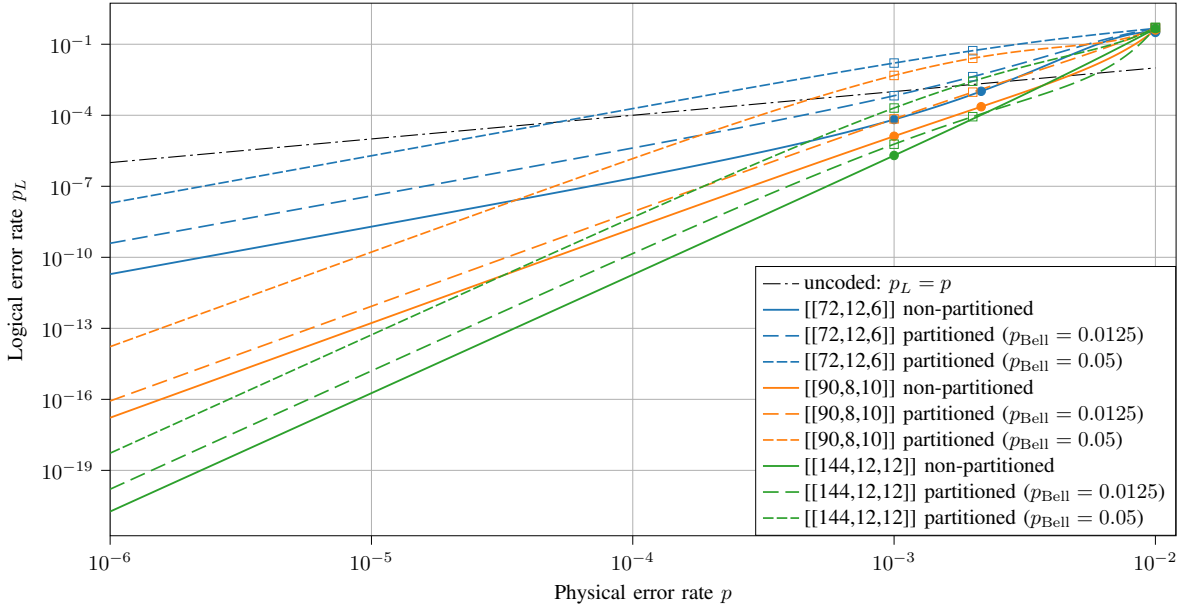}
    }
    \caption{Logical error-rate comparison between non-partitioned and partitioned BB code implementations for the $[[72,12,6]]$, $[[90,8,10]]$, and $[[144,12,12]]$ codes under circuit-level noise. Solid curves correspond to non-partitioned implementations, while dashed curves correspond to partitioned implementations obtained from a \texttt{pymetis} bipartition of the combined Tanner graph. In the partitioned case, cross-partition interactions are implemented using Bell-pair-assisted bridge operations, and the effect of Bell-link quality is shown for Bell-state fidelities $F_{\mathrm{Bell}}=0.99$ and $F_{\mathrm{Bell}}=0.96$ (equivalently, $p_{\mathrm{Bell}}=0.0125$ and $p_{\mathrm{Bell}}=0.05$). The black dash-dotted line denotes the uncoded reference $p_L=p$.}
    \label{fig:bb_partitioned}
    \vspace*{-10pt}
\end{figure*}

A full syndrome cycle means one complete application of the depth-8 BB syndrome-extraction schedule: all scheduled check-data interaction layers together with the final ancilla-measurement layers that complete that round. If a partition produces $E_{\mathrm{cut}}$ bridge edges, then one such cycle requires $E_{\mathrm{cut}}$ Bell pairs. If the circuit is repeated for $N_{\mathrm{rep}}$ syndrome cycles, then the Bell-pair demand per full circuit shot is
\[
N_{\mathrm{Bell/shot}}=E_{\mathrm{cut}}\,N_{\mathrm{rep}}.
\]
This gives a direct network interpretation of the partition quality: fewer cut edges imply fewer Bell pairs consumed per syndrome cycle (see Tables~\ref{tab:bb72_partition_stats} and~\ref{tab:bb_partition_summary}).

% The Bell-pair quality is modeled independently through a two-qubit depolarizing parameter $p_{\mathrm{Bell}}$ applied to the Bell-pair resource used in each bridge gadget.
The noise model remains the same as before for idle qubits, measurements, single-qubit gates, and CNOT operations. 
For the noise affecting the qubits sitting apart from each other, in the inter-partition bridge, we consider two-qubit depolarizing noise on the Bell pairs, characterized by an error probability $p_{\mathrm{Bell}}$ \cite{Jacinto_2026}. 
Under this convention, the corresponding noisy Bell-pair fidelity is
% $F\!\left(\rho,\lvert\Phi^+\rangle\!\langle\Phi^+\rvert\right)
% = 1 - \frac{4}{5}p_{\mathrm{Bell}}.$ 
$F(\rho, \ket{\Phi^+} \bra{\Phi^+}) = 1 - \frac{4}{5}p_{\mathrm{Bell}}$, where $\ket{\Phi^+} = \frac{1}{\sqrt{2}} (\ket{00} + \ket{11})$ and $\rho$ is the depolarized $\ket{\Phi^+}$.
Equivalently,
\[
p_{\mathrm{Bell}} = \frac{5}{4}(1-F).
\]
Thus, for example, $F=0.96$ corresponds to $p_{\mathrm{Bell}}=0.05$, while $F=0.99$ corresponds to $p_{\mathrm{Bell}}=0.0125$.

% \begin{figure}[!t]
%     \centering
%     \includegraphics[trim={0.3cm 0cm 0cm 0cm}, clip, 
%     width=0.5\textwidth]{%
% \input{bb.tikz}
% }
%     \caption {Logical error-rate comparison between non-partitioned and partitioned BB code implementations for the $[[72,12,6]]$, $[[90,8,10]]$, and $[[144,12,12]]$ codes under circuit-level noise. Solid curves correspond to non-partitioned implementations, while dashed curves correspond to partitioned implementations obtained from a \texttt{pymetis} bipartition of the combined Tanner graph. In the partitioned case, cross-partition interactions are implemented using Bell-pair-assisted bridge operations, and the effect of Bell-link quality is shown for Bell-state fidelities $F_{\mathrm{Bell}}=0.99$ and $F_{\mathrm{Bell}}=0.96$ (equivalently, $p_{\mathrm{Bell}}=0.0125$ and $p_{\mathrm{Bell}}=0.05$). The black dash-dotted line denotes the uncoded reference $p_L=p$.}
%     \label{fig:bb_partitioned}
%     \vspace*{-10pt}
% \end{figure}

All simulations use a circuit-level noise model. Local one- and two-qubit operations are followed by depolarizing noise, and the ancilla measurements inside each syndrome cycle are noisy. After the final noisy syndrome cycle, all data qubits are measured transversally to close the last detector layer.
For the non-partitioned baseline, we use the same depth-8 BB circuit but without Bell-bridge replacements. Monte Carlo sampling was performed using \texttt{sinter} with four workers, and each task was run until either $10^6$ shots or $10^3$ logical failures were reached.
To compare the partitioned and non-partitioned cases at lower physical error rates, we fit the logical error-rate curves using the ansatz \cite{Bravyi_2024}
\[
p_L(p)=p^{\alpha}\exp(c_0+c_1p+c_2p^2) \ \ ,
\ \ 
\alpha=\left\lceil \frac{d'_{\mathrm{circ}}}{2}\right\rceil,
\]
where $d'_{\mathrm{circ}}$ is the effective circuit distance used for the corresponding code family~\cite[Table 1]{Bravyi_2024}. When only two simulated data points are available for a given curve, we use the corresponding first-order truncation by setting $c_2=0$.
\begin{table}[t]
\centering
\footnotesize
\caption{Fitted coefficients for the logical error-rate ansatz
$p_L = p^{\alpha}\exp(c_0 + c_1 p + c_2 p^2)$
for the selected BB-code families. }
% For non-partitioned curves, ``--'' indicates that $p_{\mathrm{Bell}}$ is not applicable.}
\label{tab:bb_fit_coeffs}
\resizebox{\columnwidth}{!}{%
\begin{tabular}{|l|l|c|c|c|c|c|c|}
\hline
Code & Curve & $p_{\mathrm{Bell}}$ & $\alpha$ & $N_{\mathrm{pts}}$ & $c_0$ & $c_1$ & $c_2$ \\
\hline
% $[[72,12,6]]$   & non-partitioned & --     & 3 & 3 & 10.653 & 471.645   & $-2.709\times 10^4$ \\
$[[72,12,6]]$   & partitioned     & 0.0125 & 3 & 3 & 13.680 & -290.350  & $ 2.209\times 10^4$ \\
$[[72,12,6]]$   & partitioned     & 0.05   & 3 & 3 & 17.590 & -1059.125 & $ 6.059\times 10^4$ \\
\hline
% $[[90,8,10]]$   & non-partitioned & --     & 4 & 3 & 16.635 & -291.959  & $ 3.789\times 10^4$ \\
$[[90,8,10]]$   & partitioned     & 0.0125 & 4 & 3 & 18.263 & -214.952  & $ 1.572\times 10^4$ \\
$[[90,8,10]]$   & partitioned     & 0.05   & 4 & 3 & 23.547 & -1328.447 & $ 7.466\times 10^4$ \\
\hline
% $[[144,12,12]]$ & non-partitioned & --     & 5 & 2 & 21.324 & 92.262    & 0 \\
$[[144,12,12]]$ & partitioned     & 0.0125 & 5 & 3 & 23.500 & -1081.297 & $ 9.657\times 10^4$ \\
$[[144,12,12]]$ & partitioned     & 0.05   & 5 & 3 & 27.004 & -1029.260 & $ 5.650\times 10^4$ \\
\hline
\end{tabular}%
}
\end{table}

Figure~\ref{fig:bb_partitioned} compares the logical error rate $p_L$ as a function of the physical error rate $p$ for three Bivariate Bicycle (BB) codes $[[72,12,6]]$, $[[90,8,10]]$, and $[[144,12,12]]$ (from~\cite{Bravyi_2024}) evaluated under non-partitioned and partitioned configurations, with the latter tested at two Bell-pair error rates, $p_{\text{Bell}} = 0.0125$ and $p_{\text{Bell}} = 0.05$. 
In the non-partitioned case (solid curves), all three codes exhibit strong logical error suppression well below the uncoded baseline $p_L = p$, with the $[[144,12,12]]$ code achieving the lowest logical error rate of approximately $10^{-19}$ at $p = 10^{-6}$, and the slope of each curve scaling consistently with its code distance $d$. 
Partitioning, however, introduces Bell-pair noise that degrades performance in a manner strongly dependent on $p_{\text{Bell}}$: 
at $p_{\text{Bell}} = 0.0125$ the logical error rate rises by several orders of magnitude yet the codes retain a clear advantage over the uncoded baseline for $p \lesssim 10^{-3}$, whereas at $p_{\text{Bell}} = 0.05$ the degradation is severe, with the partitioned curves approaching or intersecting the uncoded line near $p \approx 10^{-3}$ and thereby losing much of their benefit. 
As $p$ approaches $10^{-2}$, all curves converge toward the uncoded baseline, reflecting proximity to the code threshold. 
Overall, these results demonstrate that while higher-distance BB codes deliver substantially stronger error suppression in the sub-threshold regime, partitioning incurs a measurable cost that is tightly controlled by Bell-pair fidelity. 
The results indicate that maintaining $p_{\text{Bell}} \lesssim 0.0125$ (or $F \gtrsim 99\%$) is essential to preserve a useful logical error advantage in networked architectures and remain competitive with monolithic architectures.

There are deeper architectural questions that must be studied further.
One question is the schedule for the purification of Bell pairs to match the syndrome measurement schedule.
Another is the necessary intra-node connectivity to interface the Bell pair qubit with different check and data qubits to perform syndrome measurements.
Yet another is the scaling of this performance as the number of nodes is increased beyond two.
Such future investigations will illuminate the (hopefully relaxed) engineering requirements for networked architectures when compared to monolithic ones.
These could drive the hardware development of even better codes such as hypergraph product and lifted product QLDPC codes.

\section{Conclusion}
\label{sec:conclusion}

In summary, we first reproduced and evaluated the networked surface code architecture described by Nickerson \textit{et al.}~\cite{Nickerson:2013ecj} with \texttt{Stim}. 
This provided a useful reference point for studying networked fault-tolerant operation under circuit-level noise, and in particular gave quantitative insight into the role of the entangled resource budget, modeled in our case through noisy GHZ-state generation.
By varying the relevant network-noise parameters in that setting, we were able to study how the quality and availability of shared entanglement influence the logical performance of the networked surface code system.

Building on that foundation, we looked at a networked version of BB codes. 
Instead of deploying an entire BB code block in one processor, we explored what happens when the code is split across multiple modules, which might be more scalable. 
To make this work, we created the combined $X$-$Z$ Tanner graph for the BB code and used it to find a way to divide both data and check nodes with as few connections between modules as possible. 
This approach lets us keep local operations simple within each module, while handling the connections between modules through special bridge operations using Bell pairs.
By doing this, we could directly see how partitioning the code affects both the circuit construction and the extra communication needed from the network.
The resulting networked BB code model allows the tradeoff between Bell-pair fidelity and logical error-rate performance to be studied explicitly.
Consequently, the simulations provide a direct way to quantify the cost of a networked BB code architecture: on one hand, partitioning may enable distributed implementation across smaller processing units, while on the other hand it introduces a communication-induced penalty that appears both in the logical error rate and in the required entanglement resources. 
This tradeoff is central when assessing whether a BB code should be deployed as a monolithic block or as a distributed code across multiple networked modules.

In the future, we would like to investigate how the number of partitions and the number of inter-module interactions might affect each other, and how sensitive the network becomes to the entanglement resources.
It is unclear if the polynomials defining the BB code directly determine the optimal partition for the balanced min-cut objective.
It will also be interesting to study if logical operations are more easily realized in such networked architectures by building structured interactions between qubits on different modules, e.g., for fold-transversal gates.
This direction will help to understand if we get a better tradeoff by distributing a code into smaller modules rather than using a whole code per module while performing modular quantum computing.

\section*{Acknowledgment}
S.~S. thanks Shantom Borah, Michele Pacenti, and Oskar Novak for their valuable inputs and help while setting up the simulation.
This work was partially supported by the National Science Foundation under Grant nos. 2106189 and 2540171.
% The preferred spelling of the word ``acknowledgment'' in America is without 
% an ``e'' after the ``g''. Avoid the stilted expression ``one of us (R. B. 
% G.) thanks $\ldots$''. Instead, try ``R. B. G. thanks$\ldots$''. Put sponsor 
% acknowledgments in the unnumbered footnote on the first page.

% \section*{References}

% Please number citations consecutively within brackets \cite{b1}. The 
% sentence punctuation follows the bracket \cite{b2}. Refer simply to the reference 
% number, as in \cite{b3}---do not use ``Ref. \cite{b3}'' or ``reference \cite{b3}'' except at 
% the beginning of a sentence: ``Reference \cite{b3} was the first $\ldots$''

% Number footnotes separately in superscripts. Place the actual footnote at 
% the bottom of the column in which it was cited. Do not put footnotes in the 
% abstract or reference list. Use letters for table footnotes.

% Unless there are six authors or more give all authors' names; do not use 
% ``et al.''. Papers that have not been published, even if they have been 
% submitted for publication, should be cited as ``unpublished'' \cite{b4}. Papers 
% that have been accepted for publication should be cited as ``in press'' \cite{b5}. 
% Capitalize only the first word in a paper title, except for proper nouns and 
% element symbols.

% For papers published in translation journals, please give the English 
% citation first, followed by the original foreign-language citation \cite{b6}.
\IEEEtriggeratref{8}
\bibliographystyle{ieeetr}
\bibliography{ieee}
% \begin{thebibliography}{00}
% \bibitem{b1} G. Eason, B. Noble, and I. N. Sneddon, ``On certain integrals of Lipschitz-Hankel type involving products of Bessel functions,'' Phil. Trans. Roy. Soc. London, vol. A247, pp. 529--551, April 1955.
% \bibitem{b2} J. Clerk Maxwell, A Treatise on Electricity and Magnetism, 3rd ed., vol. 2. Oxford: Clarendon, 1892, pp.68--73.
% \bibitem{b3} I. S. Jacobs and C. P. Bean, ``Fine particles, thin films and exchange anisotropy,'' in Magnetism, vol. III, G. T. Rado and H. Suhl, Eds. New York: Academic, 1963, pp. 271--350.
% \bibitem{b4} K. Elissa, ``Title of paper if known,'' unpublished.
% \bibitem{b5} R. Nicole, ``Title of paper with only first word capitalized,'' J. Name Stand. Abbrev., in press.
% \bibitem{b6} Y. Yorozu, M. Hirano, K. Oka, and Y. Tagawa, ``Electron spectroscopy studies on magneto-optical media and plastic substrate interface,'' IEEE Transl. J. Magn. Japan, vol. 2, pp. 740--741, August 1987 [Digests 9th Annual Conf. Magnetics Japan, p. 301, 1982].
% \bibitem{b7} M. Young, The Technical Writer's Handbook. Mill Valley, CA: University Science, 1989.
% \end{thebibliography}
% \vspace{12pt}
% \color{red}
% IEEE conference templates contain guidance text for composing and formatting conference papers. Please ensure that all template text is removed from your conference paper prior to submission to the conference. Failure to remove the template text from your paper may result in your paper not being published.

\end{document}